\documentclass[twocolumn]{aastex62}

\usepackage{amsmath}

\def\be{\begin{equation}}
\def\ee{\end{equation}}
\def\ba{\begin{eqnarray}}
\def\ea{\end{eqnarray}}

\def\12{{1\over 2}}

\def\msun{M_\odot}
\def\lsun{L_\odot}

\def\etal{{\it et~al.~}}
\def\ltgt{$\; \buildrel < \over \geq \;$}
\def\gtlt{\lower.5ex\hbox{\ltgt}}
\def\ltsima{$\; \buildrel < \over \sim \;$}
\def\simlt{\lower.5ex\hbox{\ltsima}}
\def\gtsima{$\; \buildrel > \over \sim \;$}
\def\simgt{\lower.5ex\hbox{\gtsima}}

\received{...}
\revised{...}
\accepted{...}
\submitjournal{ApJ}

\shorttitle{Submillimeter signatures from SMBHs}
\shortauthors{Vasiliev \& Shchekinov}

\begin{document}

\title{SUBMILLIMETER SIGNATURES FROM GROWING SUPERMASSIVE BLACK HOLES BEFORE REIONIZATION}

\correspondingauthor{Evgenii O. Vasiliev}
\email{eugstar@mail.ru}

\author{Evgenii O. Vasiliev}
\affiliation{Southern Federal University, Rostov on Don 344090, Russia}
\affiliation{Lebedev Physical Institute, Russian Academy of Sciences, 53 Leninsky Ave., Moscow 119991}

\author{Yuri A. Shchekinov}
\affiliation{Lebedev Physical Institute, Russian Academy of Sciences, 53 Leninsky Ave., Moscow 119991}
\affiliation{Raman Research Institute, Sadashivanagar, Bengaluru 560080, Karnataka, India}

\begin{abstract}
The presence of supermassive black holes (SMBHs) with masses up to $M_\bullet\sim10^9\msun$ at redshifts $z\simeq7.5$ suggests that their seeds may have started to grow long before the reionization in ambient medium with pristine chemical composition. During their latest 500Myr episode of growing from $z\geq10$ to $z\sim7$ the black holes shine as {luminous} as $10^{11}\hbox{--}10^{12}\lsun$, with a cumulative spectrum consisting of the intrinsic continuum from hot accretion disk, nebular hydrogen and helium spectral lines and free-free continuum from gas of host halos. Here we address the question of whether such a plain spectrum would allow us to trace evolution of these growing SMBHs. {In {our} calculations we assume that host galaxies have stellar populations with masses smaller than the mass of their central black holes -- the so-called obese black hole galaxies. Within this model} we show that for a sufficiently high mass of gas in a host galaxy -- not smaller than the mass of a growing black hole, the cumulative spectrum in the far-infrared reveals a {sharp} transition from a quasi-blackbody Rayleigh-Jeans spectrum of the black hole $\propto\lambda^{-2}$ to a flat free-free nebular continuum $\lambda^{0.118}$ on longer wavelength limit. Once such a transition in the spectrum is resolved, the black hole mass can be inferred as a combination of the observed {wavelength at the transition} $\lambda_k$ and the corresponding spectral luminosity. {Possible observability of {this effect} in spectra of growing high-$z$ SMBHs and determination of their mass with the upcoming {\it JWST} and the planned space project {\it Spektr-M} is briefly discussed. }  
\end{abstract}


\section{Introduction} 

Supermassive black holes (SMBH) with masses $M_{\bullet}\sim 10^9\hbox{--}10^{10}\msun$ are recognized recently to be {present at redshifts as high as $z\simeq 6\hbox{--}7.5$ when} the Universe was $650\hbox{--}800$ Myr {young \citep{fan03,willott10,mortlock11,wu15,banad14,banad18,decarli18,izumi19} -- in total more than {150 such SMBHs are already known \citep[see, e.g.][]{fan19}}.} Understanding of their origin remains elusive -- it is unclear how massive were their seeds, how efficient was their growth rate and what was the mass reservoir for the growth. To fit the existence of the quasars J0100 + 2802 ($z=6.33$), J1120 + 0641 ($z=7.09$) and J1342 + 0928 ($z=7.54$) with SMBH masses $M_{\bullet}=1.2\times 10^{10},~2\times 10^9$ and $7.8\times 10^8\msun$ respectively, one has to assume that their masses grow as $M_{\rm \bullet}=M_{\rm \bullet,0}\exp[t/(47~{\rm Myr})]$ corresponding to the standard Eddington limit with a 10\% radiative {efficiency $\epsilon$, the Salpeter growth time $t_S=\epsilon c\sigma_T/(4\pi Gm_p)=47$ Myr} and the seed mass $M_{\rm \bullet,0}\geq 10^3\msun$ at $z\geq 40$ \citep{banad18}. In this scenario SMBHs have to begin growing even earlier than the very first stars are assumed to have appeared \citep[see discussion in][]{barkana01}. Moreover, it suggests {that the accretion is tightly tuned} to the Eddington {rate}, what seems physically unlikely \citep[see discussion in][]{haiman01,volonteri05,haiman13,madau14,alexander14}. {Lower mass black holes with $M_\bullet\sim 100\msun$ originated from Pop III stars are apparently unlikely to serve as seeds for growing SMBHs, because photoionization and photoheating from their massive progenitors strongly suppress further supply of cold mass on to the BH \citep{johnson07}, and would require even longer time for the black hole to grow. Note however, this channel for SMBH seeds is currently widely discussed \citep[see references in][]{natarajan2020}.}  One should note that this problem of presence of such enormously massive BHs in a younger than 1 Gyr universe can be to a certain extent eased when possible magnification of $z>6$ SMBHs due to gravitational lensing is accounted \citep{fan19,pacucci19,pacucci19lens}. {Recent millimeter observations of the quasar J0100 + 2802 ($z=6.33$) with the most massive BH $M_{\bullet} \sim 1.2\times 10^{10}\msun$ known at $z>6$, indicate strong lensing with a magnification factor of $\sim 450$ \citep{fujimoto19lens}. As a result, the estimate of the SMBH mass may be reduced more than an order of magnitude, though still remains exceedingly large for a 1 Gyr universe $\sim 10^9\msun$ \citep{fujimoto19lens}. The fraction of so strongly magnified quasars is fairly low accounting a very small typical angular size of such lenses, as a rule $\ll 1^{\prime\prime}$ \citep[see, e.g.~][]{pei95,bolton08,pacucci19lens}.} 

More suitable scenarios can involve {\it i)} a hierarchical assembly of pregalactic massive black holes in $\Lambda$CDM cosmology \citep{volonteri03,yoo04,madau04,volonteri05}, {\it ii)} rapid growing of massive BH with a highly super-critical rates \citep{volonter15}, involving in particular supercritical ``slim'' disk mode \citep{begelman82,paczynski82,abram88,madau14} or a ``supra-exponential'' growth of a low-mass stellar BH in a very dense environment with low-angular-momentum velocity field \citep{alexander14}, {\it iii)} a direct collapse black hole (DCBH) of intermediate {mass ($\sim 10^4\hbox{--}10^5\msun$) growing} in a gaseous halo with a low-mass stellar population \citep[{lower than the DCBH mass at the beginning of their common evolution}][]{volonteri08,natara17} -- the obese black hole galaxy (OBG) stage as termed by \citet{agarwal13}. The first two scenarios suggest a stochastic super-Eddington feeding by high-density regions in the host galaxies and rapid growth in later epochs $z_0\leq 20$. The third scenario requires {more massive than low-mass stellar} seeds present presumably in a population of {halos hosting intermediate-mass BHs} ($M_{\bullet}\simlt 10^5\msun$) grown at earlier epochs $z\simgt 20$ under a fast DCBH growth \citep[see discussion in][]{haen93,bromm03,yosida03,lodato06,visbal14a,visbal14,agarwal16,chon16,pacucci16,latif16,pacucci17,inayoshi18,maio18,wise19}. {To dicriminate between these scenarios observations of SMBHs at even higher redshifts $z>7.5$, closer to the beginning of their growth are needed. Moreover, taking into account the case of the quasar J0100 + 2802 multiwavelength observations would be of great importance to put constrains on the early phases of SMBH growth \citep[see recent discussion in~][]{fan19,fujimoto19lens,pacucci19lens}.} 

{All scenarios predict {the} presence of higher-mass BHs growing from $M_{\bullet}\sim 10^5$ to $\sim 10^9\msun$ between} redshifts $z=20$ and $z\simeq 8$. In the Eddington regime their bolometric luminosities lie in the range $L_{\rm Edd}\sim 10^{43}\hbox{--}10^{47}$ erg s$^{-1}$. { For a medium spectral resolution $R\sim 1000$ the expected flux density is of {$S_\nu\sim 0.01\hbox{--}100\mu$Jy in infrared band $\lambda \sim 1\mu$m ($\nu\sim 300$ THz) \citep[for more discussion see ][]{natara17} and $\sim $ 1$\mu$Jy to 0.1Jy in submm band $\nu\sim 300$ GHz ($\lambda \sim 0.1$mm) \citep[see, e.g.][]{valiante18}, respectively}, compared to those measured in already discovered high-redshift SMBH \citep{fan03,willott10,banad14,banad18}. As such they} can be detected by currently ongoing and upcoming instruments. Several attempts to model fluxes from growing BHs in near and mid infrared and their observational feasibility with {\it JWST} are made by \citet{pacucci16,natara17,barrow}. {However, as far as possible longer wavelength features (such as higher subordinate hydrogen line series, or free-free continuum) from even earlier phases of BH growth beyond $z>10$ are concerned, the {\it JWST} capabilities can become inapplicable for detection of spectral manifestations from low-metallicity or pristine gas. } 

In this paper we address the question of whether observational signatures from growing BH in epochs covered by redshifts between $z=20$ and $z=7.7$ can be observed in far infrared and submillimeter wavebands, and whether the growing regime can be recognized from these manifestations. We argue that besides traditional line spectroscopy,  spectral features in infrared and far-infrared continuum might be an efficient complementary tool for studying physical conditions in gas ionized and heated by a growing massive BH and the characteristics of the BH itself. 

In next section the model we use in our calculations is described: {the accretion mode, the models of the BH and stellar population emission spectra and their interrelation}, Sec. \ref{sec:ion} contains the results: {{the} evolution {of} the line and the continuum emissions from growing BHs and ionized ambient gas}, in Sec. \ref{sec:disc} we discuss issues related to observability {of growing BHs at redshifts $z>7.5$ with the planned IR telescopes}, Sec. \ref{sec:sum} summarizes the results.  

\section{Model description} \label{model}

\subsection{Growth of the black hole} \label{sec:bhgrowth}

\begin{figure}
\center
\includegraphics[width=85mm]{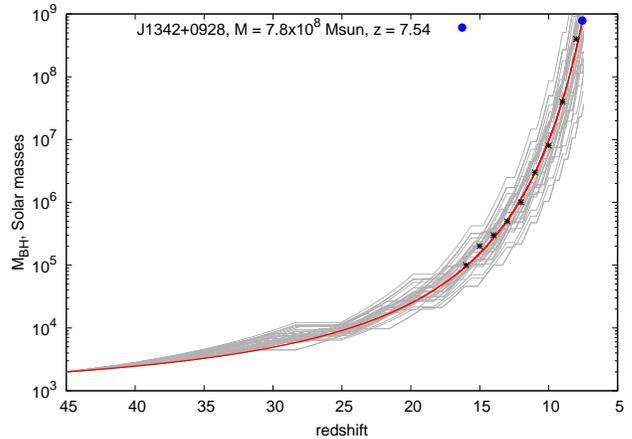}
\caption{
Black-hole growth model. Blue point (solid circle) corresponds to the quasar J1342 + 0928 ($z=7.54$) \citep{banad18}, {grey lines show the trajectories of growing BH masses from Eq. (\ref{rand}) for different sets of $t_i$, $\eta_i$ and $\epsilon_i$ constrained by the condition {$M(z=8\pm 1.5)=3\times 10^8\msun$}. Red line shows the average BH mass growth law, black dots around the red line indicate the black hole masses for which we calculate the spectra. Black points present one of the trajectories.}
}
\label{fig-mbh}
\end{figure}

{The red line in Fig.~\ref{fig-mbh} depicts an illustrative scenario with a continuous exponential growth rate $\dot M\propto M$ \citep{volonteri03,shapiro05,madau14}  with the critical (Eddington) regime} 
 
\be\label{vol}
 M(t)=M_0 {\rm exp}\left({1-\epsilon \over \epsilon} {t \over 0.47~{\rm Gyr}}\right)
\ee 
{where $M_0 = 1.6 \times 10^3~\msun$ at $z\geq 40$ and $\epsilon = 0.095$ being a mass-to-energy conversion factor \citep{soltan82,volonteri03} are assumed} for $M(t)$ to reach $M_\bullet=7.8\times 10^8\msun$ at $z=7.5$ as supposed by \cite{banad18}. Such a regime suggests a mass source {that enhances its efficiency to} a continuously tuned feeding {equivalent to} the critical accretion rate $\dot M\propto L_{\rm Edd}\propto M(t)$, where $L_{\rm Edd}$ is the Eddington luminosity. A more realistic scenario can imply {sporadic} enhancement of the accretion rate to the super-Eddington level and subsequent quiet episodes, such that when averaged over large time the accretion  would correspond to the regime resulting eventually in a $10^9\msun$ class black hole at $z\simeq 7$ \citep{madau14,pacucci17}. For these reasons we consider another model with episodically {enhanced accretion from radiatively inefficient disks with a faster growth rate $\dot M_i(z)$}
\be
 \dot M(t) =  \sum\limits_i^{}{\eta_i(1-\epsilon_i)\over\epsilon_i}{M_i\over 0.45~{\rm Gyr}}\Theta(t),
 \label{rand}
\ee
where
\begin{equation*}
 \Theta(t) = 
  \begin{cases} 
    1, & t_i\leq t\leq t_i+\Delta t_i\\
    0, & {\rm otherwise},
  \end{cases}
  \label{sample}
\end{equation*}
with $M_i=M(t_i)$ {being the mass reached during the $i$-th episode of accretion}, $\eta_i=L_i/L_{\rm Edd}\gtlt 1$ {being a random number chosen in the limits $\eta_1<\eta<\eta_2$ such to ensure the mean mass of a {grown BH $M_\bullet=3\times 10^8\msun$ at} $z=8$ which closely fits masses of observed BHs at $z=7.5$}, $\epsilon_i=\epsilon \simlt 0.095$ is assumed; $t_i$ and $\Delta t_i$ are choosen such {to match $\langle M(t)\rangle$ to $M_\bullet=3\times 10^8\msun$ at $z=8$ as depicted by red line} in  Fig.~\ref{fig-mbh}. {Further we calculate the spectra for one of the trajectories depicted in Fig.~\ref{fig-mbh} by black points.} With regard to the spectral features of growing BHs we are aiming here, one has to note that {the} models with randomly varying accretion (\ref{rand}) differ from the Eddington one (\ref{vol}) only by their observed fluxes versus the redshifts $z=z(F_\nu)$ corresponding to a given BH mass. In other words, once the redshift of a source is identified {the only relevant parameters are the spectral shape and the measured flux.}

\subsection{Spectrum of the accreting BH} \label{sec:bhspec}

\begin{figure}
\center
\includegraphics[width=85mm]{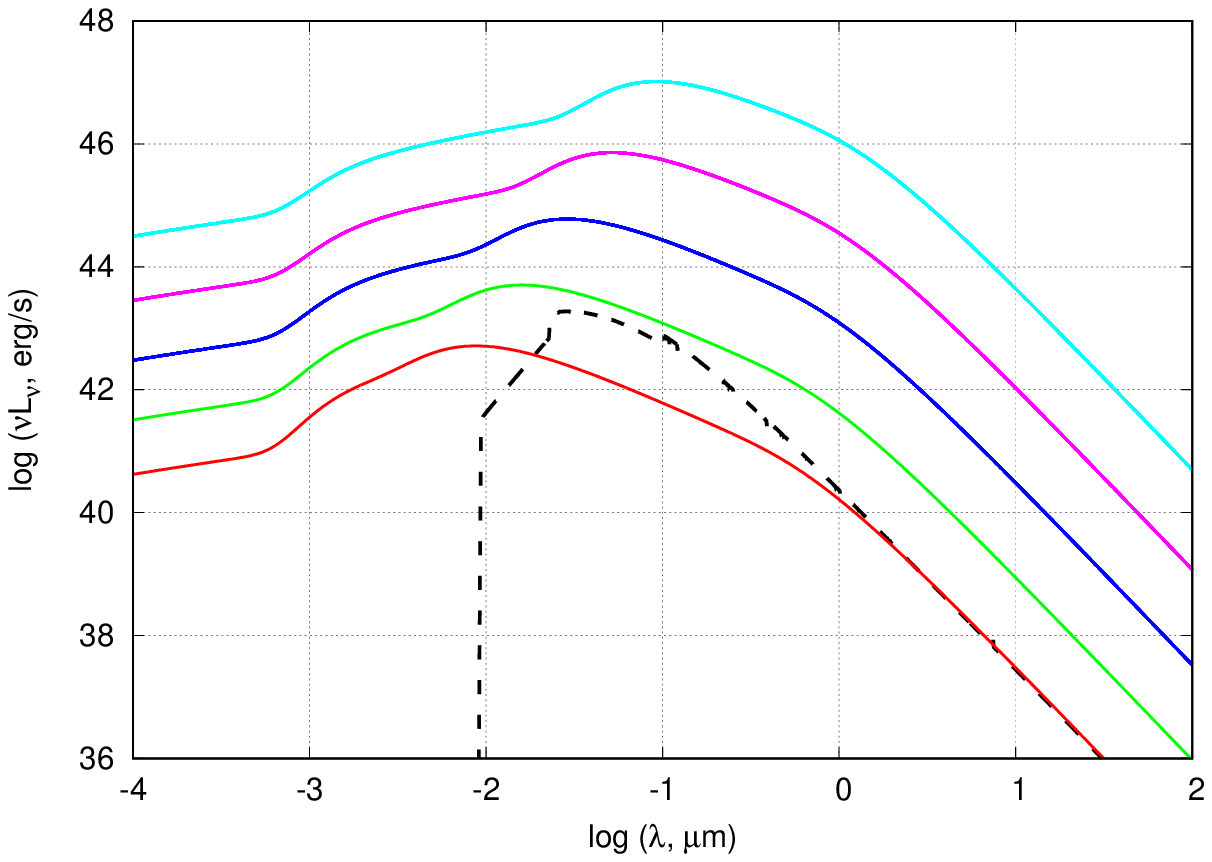}
\caption{
The SEDs for BHs with masses $10^5, 10^6, 10^7, 10^8$ and $10^9~\msun$ (solid color lines from bottom to top); clearly seen is the soft X-ray {excess at $\lambda\sim 0.01\hbox{--}0.03\mu$m ($100\hbox{--}300$ \AA\,), from} the intermediate warm comptonizing region of the accretion disk \citep{kubota18}.The SED of the stellar bulge of a host galaxy as modelled by \citet{zack11} for the metal-free stellar population PopIII.1 is shown by dashed black line, red line show spectra of central BHs with the mass $M_\bullet\sim 0.02 M_\ast$ {corresponding to the mass of stellar population $M_\ast=5\times 10^6\msun$}.
}
\label{fig-spec-kubota}
\end{figure}

In our calculations we use a broadband spectral energy distribution (SED) of active galactic nuclei described by \cite{kubota18,kubota19}. It is based on the slim disk model \citep{abram88} of a radially stratified disk with the three dominant regions: the hot inner disk extending from the innermost stable circular orbit $R_{isco}$ to loosely defined edges of the hot Comptonizing region $R_h$, the warm Comptonizing region from $R_h$ to $R_w$, and the outer region from $R_w$ to $R_{\rm out}$ with a flat  radial emissivity profile $F(r)\propto r^{-2}$ through over the disk. Within this model advection and wind outflow enhance radiation transfer and thus stabilize super-Eddington disks against radiation-driven instability typical for standard disks with an exceeding luminosity \citep{kubota18}. The warm intermediate disk region produces the soft X-ray excess \citep{kubota18,kubota19} contributing importantly into ionization and heating of the interstellar gas (see below) and consequently into the nebular spectrum. The bolometric luminosity is assumed to be equal to the Eddington luminosity corresponding to the average value of the BH mass depicted by solid red line in Fig. \ref{fig-mbh}: $L_{\rm Edd} = 1.26\times 10^{38} \langle M_\bullet(t)\rangle$~erg~s$^{-1}$, $\langle M_\bullet\rangle$ is in solar masses. Figure~\ref{fig-spec-kubota} presents {several} examples of SEDs for BHs with masses $10^5, 10^6, 10^7, 10^8, 10^9~\msun$ (from bottom to top).

\subsection{Spectrum of the stellar population} \label{sec:stel} 

For the stellar bulge SED {shown in Fig. \ref{fig-spec-kubota}} we utilized the population synthesis SED by \citet{zack11} for the metal-free composite stellar population produced by {a 30 Myr long} burst of PopIII.1 star formation (extremely top-heavy IMF: $50-500\msun$, {the} Salpeter slope) with a constant rate\footnote{http://www.astro.uu.se/~ez/yggdrasil/yggdrasil.html}. It is worth stressing that such an extreme IMF has the highest contribution of stellar population into the domain of high-energy photons -- X-ray and EUV, as compared to other models: for instance, a Pop II with metallicity ${\rm [Z/H]}\sim -2$ and a Kroupa  IMF with $M_{\rm min}=0.1\msun$ and $M_{\rm max}=100\msun$, or a log-normal IMF around $M\sim 10\msun$ with $\sigma_M\sim 1\msun$ and with wide wings extending to 500 $\msun$ \citep{reiter10}. However, the stellar contribution into ionizing (X-ray and EUV) photons can be neglected if $M_\bullet\sim 0.02M_\ast$.

\subsection{Cumulative incident spectrum} \label{sec:cumul} 

{As inferred for supermassive black holes in the local Universe their masses correlate with the stellar bulge of host galaxies \citep[see reviews in][]{marconi03,haring04,sani11,kormen13,heckman14}}. {In spite} of a high spread \citep[an order of magnitude, see e.g.,][]{kormen13,heckman14} an approximate proportionality $M_\bullet/M_\ast\sim 0.002$ \citep[see Fig.~18b in][]{kormen13,decarli18} can be loosely accepted for low-redshift host galaxies. Observations of the [CII] 158 $\mu$m line in a set of $z\simgt 6$ quasars lead \citet{walter04,decarli18} to conclude that the ratio $M_\bullet/M_\ast$ in SMBHs hosting galaxies at $z\simeq 6-7$ epoch is an order of magnitude higher than in the local Universe: $\langle M_\bullet/M_\ast\rangle\sim 0.02$ at $z\sim 7$ versus 0.002 at $z\sim 0$, in conflict with a common scenario of coeval evolution of the stellar population and the central massive BH. This circumstance may either reflect a more efficient growth of BHs as compared to a possibly quenched star formation, or correspond to the above mentioned OBG stages \citep[][]{agarwal13,agarwal16} with a relatively low stellar mass in hosts galaxies. In concord with this we assume that a dominant fraction of baryons of host galaxies is in the form of gaseous halo with the mass $M_b = M_\bullet/\alpha$, with $\alpha = 0.002$ as a fiducial value, thus making the gas a sufficient reservoir for formation of stars at later epochs. The gas is assumed to occupy homogeneously a spherical layer with {the} density $n_l = 1$~cm$^{-3}$, which is typical for a diffuse ISM in a {virialized halo at $z\sim 20$ \citep[e.g.,][]{barkana01}}. Thus, the spherical layer has thickness equal to $\Delta R\sim M_b/(4\pi R^2m_p n_l)$, where $m_p$ is the proton mass, $R$, the radius of a halo. The inner radius is kept fixed $r_{in}=30$~pc. In what follows we will restrict the mass of stellar population by the ratio $M_\bullet/M_\ast\simgt 1$. 

As mentioned above for such a high $M_\bullet/M_\ast$ the BH with Eddington {luminosity $L_{\bullet,bol}\geq 1.3\times 10^{38}(M_\bullet/\msun)$ obviously dominates in the whole energy range from high energy bands (X-ray, EUV) to optical and near and partly mid infrared waves. {Indeed,} for stellar mass-to-luminosity ratio $M_\ast/L_\ast\sim 0.01\msun/\lsun$ typical for a top-heavy PopIII IMF and {even} for $M_\bullet/M_\ast=0.02$ one estimates $L_{\bullet,bol}\simgt 6L_\ast$. The far infrared is dominated by stellar population, however its overall energetics is negligible.} Therefore, under such conditions thermal and ionization state of the ISM of a host galaxy is totally determined by radiation of the central growing BH. From this point of view the soft X-ray excess along with EUV photons from the intermediate accretion disk \citep{kubota18} plays {a crucially important} role.  	 

\begin{figure*}
\center
\includegraphics[width=120mm]{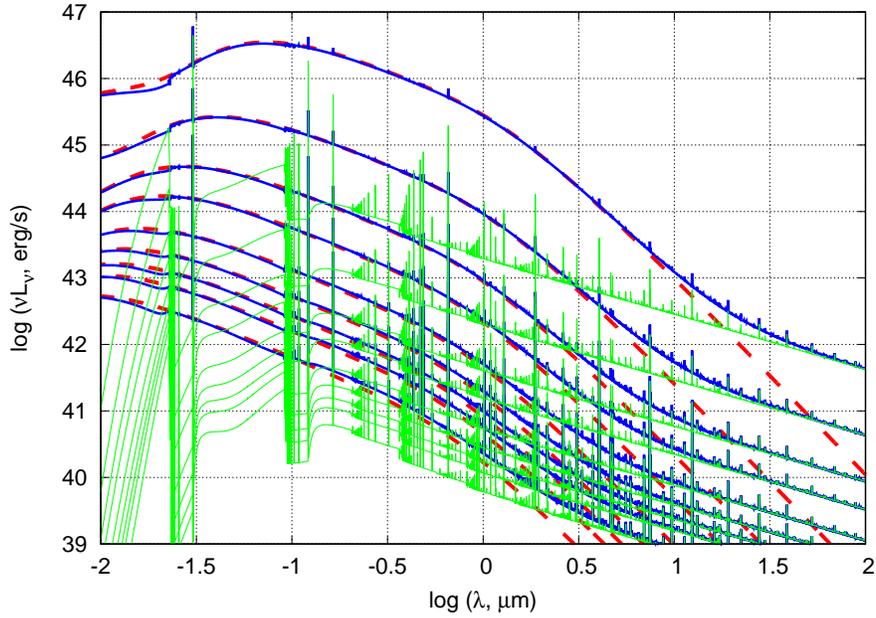}
\caption{
The incident continuum (red dashed lines), the net transmitted continuum, i.e. the sum of the attenuated incident and nebular continua along with spectral lines (blue thick lines), the nebular emission of hot gas in continuum and lines (green thin line) for the BH masses evolved following to Figure~\ref{fig-mbh} for $z = 16$ to 8 with $\Delta z=1$ (from bottom to top).
}
\label{fig-spec-bh}
\end{figure*}

\section{Results} 

\subsection{Nebular emission spectra} \label{sec:ion}

Photons emitted by an accreting BH are transmitted through the surrounding gaseous layer. In the layer we assume {photoionization and thermal equilibrium.} We use CLOUDY \citep[ver.~17,][]{cloudy17} to obtain the ionization composition and the transmitted spectrum. Figure~\ref{fig-spec-bh} shows the incident continuum radiation (red lines), {the nebular emission from the ionized gaseous layer in spectral lines and continuum} (green lines), and the sum of the two (blue lines) for the BH masses evolved following Figure~\ref{fig-mbh} for $z = 16$ to 8 with a step $\Delta z=1$. 

\subsubsection{Line emission} 

The strongest lines in the spectra presented in Figure~\ref{fig-spec-bh} are {the longest wavelength lines} of the hydrogen series: Ly$\alpha$ (1215\AA), H$\alpha$ (6563\AA), Pa$\alpha$ (1.875$\mu$m), Br$\alpha$ (4.05$\mu$m), Pf$\alpha$ (7.46$\mu$m), Hu$\alpha$ (12.37$\mu$m), H$\beta$, H$\gamma$, HeII~1640\AA. HeII~1.01$\mu$m are also  worth mentioning. The lines are more clearly seen above the continuum for less massive BHs, although they remain recognizable (except only HeII~1.01$\mu$m line) even for the most massive BHs considered here. These lines can be used not only for measuring redshifts of BH host galaxies, but also for identification the source as an OBG  candidate. Their luminosities in optical lines ($\lambda<1\mu$m) exceed $10^{42}$ erg~s$^{-1}$ for redshifts $z\simlt 12$, and the thermal widths range in $25\hbox{--}35$ km~s$^{-1}$. Their fluxes in H$\alpha$ line can reach up to several $\mu$Jy and as such may be detected even in the high resolution spectral mode of the Near-Infrared Spectrograph\footnote{http://sci.esa.int/jwst/45694-nirspec-the-near-infrared-spectrograph/} installed on {\sl JWST} \citep[see e.g.][]{kalirai}. {The IR lines may be resolved for massive BH grown in earlier epochs $z \simgt 9$ with line luminosities as high as $10^{42}$erg~s$^{-1}$. Indeed, for example, the expected flux in Pf$\alpha$ (7.46$\mu$m) line being of order $\sim 30\mu$Jy, is higher than the detection limit for moderate spectral mode ($R=1000$) of the planned space telescope {\it Spektr-M} \citep[{\it Millimetron}, see in][]{kardash14}.} On the other hand the flux from even the most luminous BHs with line luminosity $\sim 10^{43}$erg~s$^{-1}$ at $z\sim 8$ presented in Figure~\ref{fig-spec-bh} is close to the {low-resolution ($R=300$) sensitivity threshold of} SAFARI spectrograph  \citep[being a part of the {\sl SPICA} mission, see][]{safari,spica}.

\subsubsection{Continuum} \label{sec:con}

Numerical models of radially stratified accretion slim disks with a shallow radial luminosity profiles reveal a slightly weaker dependence of the effective disk temperature on the BH mass \citep{kubota19}, than the one predicted in the analytical Novikov-Thorne disk emissivity model. It is seen from Fig. \ref{fig-spec-kubota} where the intrinsic spectrum of a BH in the long wavelength limit $\lambda>1~\mu$m behaves as a black-body with the effective temperature {$T_{eff}\propto M_\bullet^{1.5}$ }

\be \label{lbh}
{\cal L}_\nu^\bullet \equiv \nu L_{\nu,\bullet} \simeq 6\times 10^{46}M_{\bullet,9}^{1.5}\lambda_1^{-3}~{\rm erg~s^{-1}}, ~\lambda\geq 1~\mu{\rm m}, 
\ee 
versus $T_{eff}\propto M_{\bullet}^{1.8}$ for Novikov-Thorne emissivity \citep{novikov73}, here $M_{\bullet,9}=M_{\bullet}/10^9\msun$, $\lambda_1=\lambda/1 {\mu{\rm m}}$. 

Ionized and heated ISM gas emits a considerable fraction of the BH energy in free-free (bremsstrahlung) continuum. In the limit of long wavelengths $\lambda\geq 0.3~\mu$m it is \citep[see in][]{draine11} 
\be \label{lff}
{\cal L}_\nu^{ff} \equiv \nu L_\nu^{ff}\simeq 3.4\times 10^{41} T_4^{-0.323} n_e M_{g,9}\lambda_1^{-0.882}~{\rm erg~s^{-1}}, 
\ee 
where the fractional ionization $x$ in the emitting region is taken $x\simeq 1$, $M_{g,9}=M_g/10^9\msun$ is gas mass, {$T_4=T/10^4~{\rm ~K}$, is gas temperature}. 

When the interrelation between the gas and BH masses $M_g=500M_{\bullet}$ is explicitly assumed (as for the case shown in Fig. \ref{fig-spec-bh}) one can find {that the free-free nebular continuum overshines} the intrinsic radiation from the BH at the wavelength $\lambda\geq \lambda_{1,k}\simeq 0.1M_{\bullet}^{0.26}$~$\mu$m and the spectrum changes its slope -- {the 'kink' in the spectrum}. In the example shown in Fig. \ref{fig-spec-bh} it gives for the wavelength of this ``kink'' $\lambda_{k}\simeq 15~\mu$m for the SMBH with $M_{\bullet}=4\times 10^8\msun$ at $z=8$. The corresponding luminosity at the kink is ${\cal L}_k \sim 5\times 10^{42}$ erg~s$^{-1}$, as seen from Fig. \ref{fig-spec-bh}; for an arbitrary $z$ within this model:  ${\cal L}_k\propto 5.6\times 10^{35} M_{\bullet}^{0.76}$  erg~s$^{-1}$, where $M_\bullet$ in solar masses. 

The ``kink'' transition can be detected in continuum of BHs with $M_{\bullet} \simgt 10^8\msun$ at $z\simlt 9$ by future instruments since the photometric flux $\nu F_\nu = (1+z) {\cal L}_k/(4\pi d_L^2) \simeq 8\times 10^{-28} M_\bullet^{0.86}$~W~m$^{-2}$, where $d_L$ is the luminosity distance. For instance, the flux limit for imaging mode of low-resolution SAFARI spectrometer \citep{safari,spica}, is around $10^{-20}$W~m$^{-2}$ for a 10~hr integration time at wavelength $\sim 100~\mu$m, where the ``kink'' takes place. {However,} spectral observations ($R=100$) are less optimistic for the SPICA mission: $F_\nu \simeq 2.5\times 10^{-12} \lambda_{1,obs}^{5.8}~{\rm \mu Jy}$, where $\lambda_{1,obs} = (1+z)\lambda_{1,k}$ is the observed 'kink' wavelength in ${\rm \mu m}$. For instance, the maximum flux $\sim 10~\mu$Jy  reached for $M_{\bullet} \sim 8\times 10^8\msun$ at $z=8$ is about one and a half order of magnitude lower than the limit detected by SAFARI spectrometer within 10~hr integration time.

In {general} case when the gas mass is a free parameter not connected to the BH mass, the transition from {the} {bremsstrahlung {with slope} $\beta=0.118$ to {the} quasi-blackbody {spectrum with} $\beta=2$} occurs at $\lambda_k$ depending on gas mass $M_g$. {The two observables: the ``kink'' wavelength $\lambda_k$ and the corresponding luminosity  ${\cal L}_k\simeq 2{\cal L}_{\nu_k}^\bullet$ -- allow to derive the two different variables: $M_\bullet$ and $M_g$. The first follows immediately from Eq (\ref{lbh}), while the second can be derived from Eq (\ref{lff}) with accounting that ${\cal L}_k\simeq 2{\cal L}_{\nu_k}^{ff}$ and that $\lambda_k$ is linked to ${\cal L}_k$ and $M_\bullet$ from Eq (\ref{lbh}). Eventually we arrive at the following relation} 
\be 
M_{\bullet}\sim {\cal L}_{k,47}^{0.68} \lambda_k^2, 
\ee
and 
\be 
M_g\sim 2\times 10^5 n_e T_4^{-0.323} {\cal L}_{k,47}^{0.68} \lambda_k^{0.9},
\ee
correspondingly, all masses are given in $10^9\msun$, ${\cal L}_{k,47} = {\cal L}_k/10^{47}$~erg~s$^{-1}$. 

\begin{figure}
\center
\includegraphics[width=85mm]{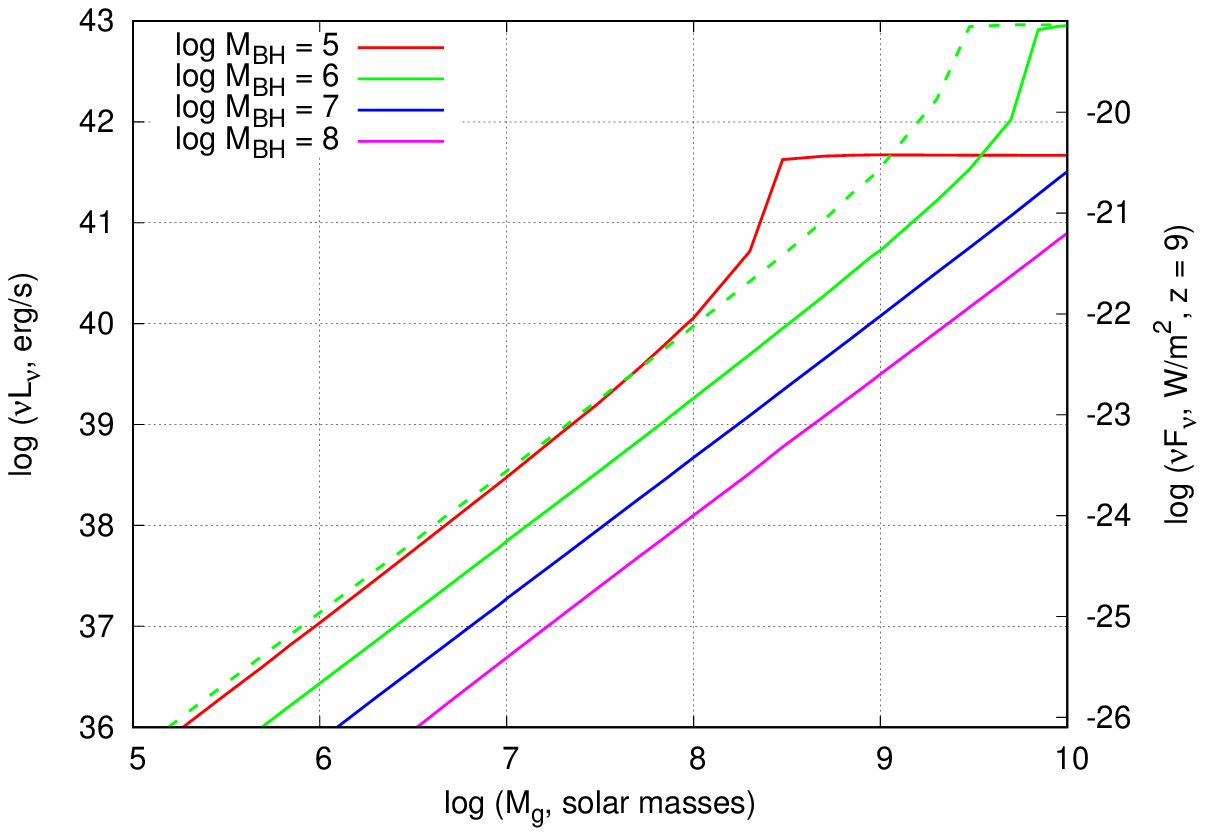}
\includegraphics[width=85mm]{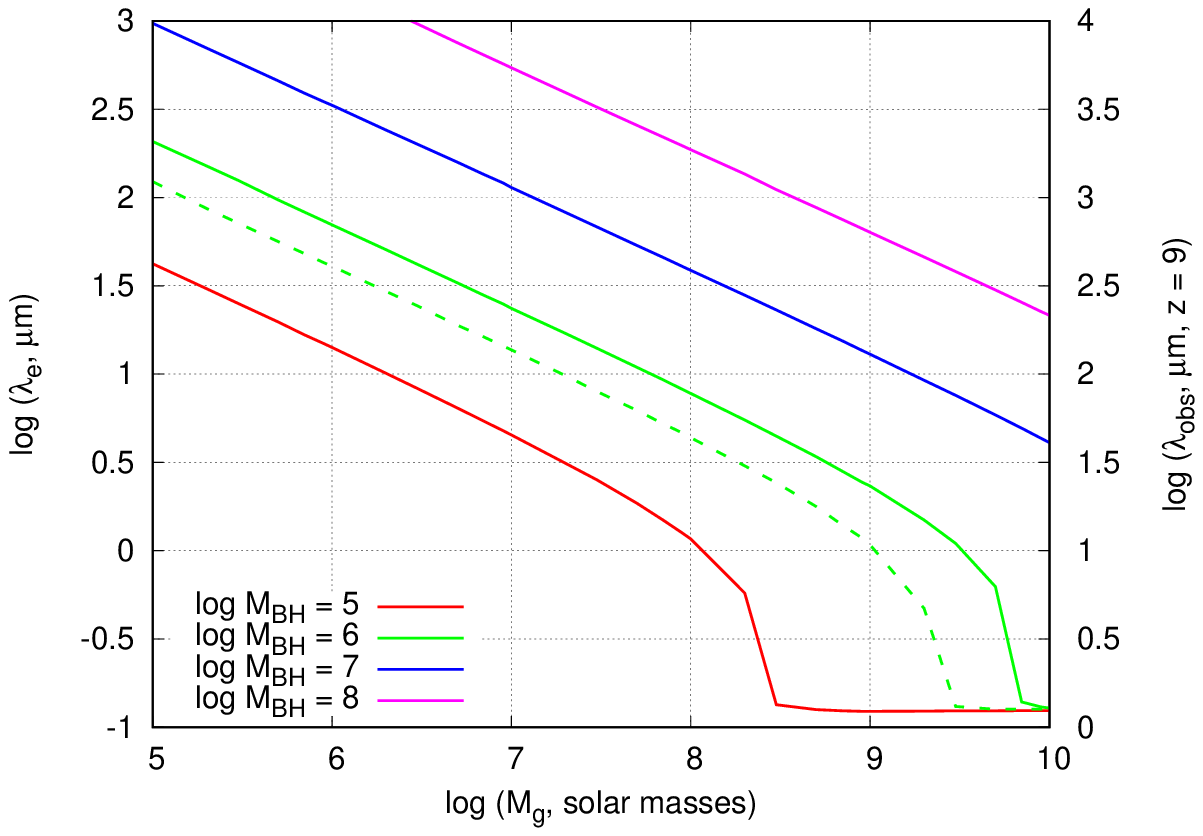}
\caption{
Luminosity at the ``kink'' ${\cal L}_k$ (upper panel) and the ``kink'' wavelength $\lambda_k$ (lower panel) vs. mass of the emitting gas layer for several values of BH mass $M_{\bullet}$ (color lines). Solid lines show the models for $n=1$~cm$^{-3}$, dashed line depicts the model for $n=3$~cm$^{-3}$ and $M_{\bullet} = 10^6\msun$: it is seen that at a fixed gas mass ${\cal L}_k$ and $\lambda_k$ scale with gas density approximately as $\sim n^{0.4}$ and $\sim n^{-0.47}$, correspondingly, concordant to arguments in Sec. \ref{sec:con}. Right axes correspond to the flux and the wavelength emitted at redshift $z=9$. 
}
\label{fig-lum-ipoint}
\end{figure}

\begin{figure}
\center
\includegraphics[width=85mm]{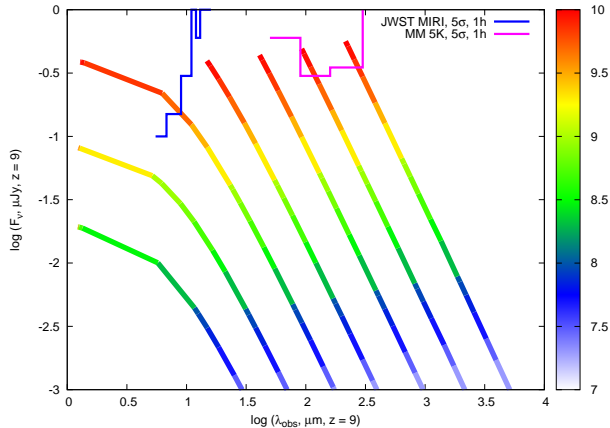}
\caption{
The wide-band (the spectral resolution $R=100$) flux of free-free continuum at the kink $\lambda_k$ emitted at redshift $z=9$ for several values of BH mass $M_{\bullet}$: $10^5,~10^{5.5},~10^6,~10^{6.5},~10^7,~10^{7.5},~10^8\msun$ from left to right; color displays the gas mass as shown {in color bar}. Blue and pink lines depict the expected detection limits (1 hour integration) of {\it JWST MIRI} and {\it Spektr-M} (MM) telescopes, correspondingly. 
}
\label{fig-flx-ipoint}
\end{figure}

\section{Discussion} \label{sec:disc}

As mentioned above an approximate proportionality between a central BH mass and stellar mass of galaxies hosting BHs in the local universe  $M_\bullet/M_\ast\sim 0.002$ \citep[e.g.][]{kormen13,heckman14,decarli18} tends to increase by an order of magnitude to higher redshifts \citep{walter04,decarli18}. It might indicate that a coeval interrelation between $M_\bullet$ and $M_\star$ was slightly shifted towards BHs in the earlier epochs, in the spirit of obese black hole galaxies, such that the galaxies' gas components in BH host galaxies served as a reservoir for feeding BHs and forming stars. Within this assumption one may think that the gas mass in a host galaxy is one of the major factors determined relation $M_\bullet/M_\star$ in the course of evolution. Our estimates given above are done within this toy scenario. 

Figure~\ref{fig-lum-ipoint} presents the luminosity in continuum $L_k$ (upper panel) and the wavelength $\lambda_k$ (lower panel) at the ``kink'', where the free-free nebular emission overshines the BH quasi-blackbody versus the gas mass $M_g$ exposed to a BH of fixed mass $M_{\bullet}$. Increase of the gas mass for a fixed $M_{\bullet}$ results in a proportional increase of the nebular emission. The ``kink'' wavelength $\lambda_k$ is shifted shortward for a higher gas mass. The flat parts in $L_k$ and $\lambda_k$ curves correspond to a saturation of luminosity when the ``kink'' transition wavelength becomes close to the Lyman break, i.e. the nebular emission overshines the BH at wavelengths $\sim 1000$\AA \ (see red and green lines in Figure~\ref{fig-lum-ipoint}). {Increase of gas density manifests in an increase of free-free continuum and a shortward shift of the ``kink'' transition wavelength (see green dashed and solid lines in Figure~\ref{fig-lum-ipoint}).} Increase of a BH mass for the fixed gas mass leads to a shift of the ``kink'' transition wavelength longward where the free-free continuum is lower.

{Interrelations between expected fluxes and wavelengths shown in Figure~\ref{fig-lum-ipoint} when the emitting objects are located at $z=9$, are combined into Figure~\ref{fig-flx-ipoint}. Even for massive BHs $M_\bullet\sim 10^8\msun$ the flux is less than $\sim 1~\mu$Jy}. Such a flux lies at the sensitivity threshold for imaging mode ($R=3$, 1 hour integration) of the planned space telescope {\it Spektr-M} \citep[see in][]{kardash14}. {For the SAFARI spectrometer this flix is about 100 times below the limit. Note that} the flux varies with gas mass as $\nu F_\nu \sim M_g^{1.5}$, while weakly sensitive to its density.

{The critical} issue for the models considered here is the observability of SMBHs grown from DCBH seeds on the OBG phase. Besides the question of their brightness the key issue is how many such objects can be met in the field of view (FoV) of a telescope. Recent theoretical models predict the number denisty of DCBH {at redshifts $z \sim 10\hbox{--}13$ in the range} $n(z) \sim 10^{-7} - 3\times 10^{-6}$ per comoving Mpc$^3$ \citep{dijkstra14,wise19}. The SMBH mass function is expected to have a peak around $\sim 10^5\msun$ \citep{basu19}. The most plausible SMBH fraction with $M_\bullet \sim 10^7-10^8\msun$ is $\simgt 0.03\hbox{--}0.1$ for an Eddington or supercritical growth regime \citep{basu19}. Thus, the number of SMBH with $M_\bullet \sim 10^7-10^8\msun$ in the redshift range $z\sim 10-11$ is $N \sim 0.03 - 2$ objects per square arcmin. Thus, one would require less than 5 random pointings {of the {\sl JWST} FoV ($2'\times 2'$) for} detection at least one of such objects. For the planning {\it Spektr-M} project with a $6\times 6'$ FoV a handful of such objects can be met in even one pointing.

\section{Summary} \label{sec:sum}

In this paper we calculated spectral features of growing massive black holes on the stages when the host galaxy stellar population is underdeveloped with a mass not exceeding the black hole mass. We assumed that a black hole begins growing from a low-mass seed at early epochs {($z\simgt 20$). The feeding rate is assumed to be kept on average in the Eddington} accretion regime in order to increase its mass upto $M_{\bullet}\sim 3\times 10^8\hbox{--}10^9\msun$ to the redshifts $z\simeq 7.5$ as observed \citep[see, e.g, in][]{mortlock11,banad18,decarli18}. We showed that 
\begin{enumerate} 
\item While growing the black hole spends a considerable fraction of its hard photons of X-ray and EUV bands to ionize and heat interstellar gas of the host galaxy. 
\item Interstellar gas re-radiates the ionizing photons of the BH in EUV, optical and infrared bands in continuum and line emission with intensities depending on the BH growing rate, and thus can serve for diagnostic of its evolutionary stages.  
\item At longer wavelengths -- in infrared and far-infrared bands, bremsstrahlung continuum re-radiated by the ISM gas overshines the continuum from the BH resulting in a change of spectral index from a Rayleigh-Jeans like $\propto\lambda^{-2}$ at shorter wavelengths to the flat free-free $\lambda^{0.118}$ in far-infrared. The wavelength corresponding to such a transition at $\lambda\sim\lambda_{k}$ and the luminosity ${\cal L}_k$ can trace the BH evolutionary stage: the BH mass can be inferred as $M_\bullet(z)\propto{\cal L}_k^{0.68}\lambda_k^2(z)$, the mass of the free-free emitting ISM gas is $M_g\propto{\cal L}_{k}^{0.68} \lambda_k^{0.9}$.    
\end{enumerate} 

\vspace{1cm}

{We thank the referee for careful reading and helpful suggestions. We thank also I. Khrykin for introducing us into pythoning, and T. Larchenkova for discussing the effects of  gravitational lensing}. This work is supported by the joint RFBR-DST project (RFBR 17-52-45063, DST P-276). EV is grateful to the Ministry for Education and Science of the Russian Federation (grant 3.858.2017/4.6). The work of YS is done under partial support from the joint RFBR-DST project (17-52-45053), by the project 01-2018 ``New Scientific Groups LPI'', and the Program of the Presidium of RAS (project code 28).


\end{document}